\documentclass[10pt]{IEEEtran}

\usepackage{amsmath,amsopn,amssymb,bm}
\usepackage{amsfonts}
\usepackage{amsthm} 
\usepackage{graphicx,xspace,color}
\usepackage{epsfig}
\usepackage{longtable,multirow}
\usepackage{array}
\usepackage{stfloats}
\usepackage{cuted}
\usepackage{cite}
\usepackage[nolist]{acronym}
\usepackage{soul}
\usepackage{algorithm}
\usepackage[noend]{algpseudocode}
\usepackage{algcompatible}
\usepackage{enumerate}
\usepackage{lettrine}
\usepackage{mathtools}
\usepackage{subcaption}
\usepackage{xcolor}
\usepackage{relsize}

\newtheorem{definition}{Definition}

\usepackage[font=footnotesize]{caption}

\graphicspath{ {./figures/}}

\begin{document}

\begin{acronym}
	\acro{RU}{resource unit}
	\acro{RUs}{resource units}
	\acro{OFDMA}{orthogonal frequency division multiple access}
	\acro{OFDM}{Orthogonal frequency division multiplexing}
	\acro{SRA}{scheduling and resource allocation}
	\acro{WLANs}{wireless local area networks}
	\acro{AP}{access point}
	\acro{STA}{station}
	\acro{STAs}{stations}
	\acro{MCS}{modulation and coding scheme}
	\acro{UL}{uplink}
	\acro{DL}{downlink}
	\acro{TF}{trigger frame}
	\acro{TFs}{trigger frames}
	\acro{SNR}{signal-to-noise ratio}
	\acro{PF}{proportional fairness}
	\acro{TWT}{target wake time}
	\acro{IoT}{Internet of Things}
	\acro{nRT}{non-real-time}
	\acro{CDF}{cumulative distribution function}
\end{acronym}

\title{A Scheduling Policy for Downlink OFDMA in IEEE 802.11ax with Throughput Constraints}

\author{Konstantinos Dovelos and Boris Bellalta 

\thanks{The authors are with the Department
	of Information and Communication Technologies, Pompeu Fabra University, Barcelona, Spain (e-mail: konstantinos.dovelos@upf.edu, boris.bellalta@upf.edu). This work has been partially
	supported by CISCO (CG\#890107-SVCF), WINDMAL PGC2018-
	099959-B-I00 (MCIU/AEI/FEDER,UE), and SGR-2017-1188.}
}

\maketitle

\begin{abstract}
In order to meet the ever-increasing demand for high throughput in WiFi networks, the IEEE 802.11ax (11ax) standard introduces orthogonal frequency division multiple access (OFDMA). In this letter, we address the station-resource unit scheduling problem in downlink OFDMA of 11ax subject to minimum throughput requirements. To deal with the infeasible instances of the constrained problem, we propose a novel scheduling policy based on weighted max-min fairness, which maximizes the minimum fraction between the achievable and minimum required throughputs. Thus, the proposed policy has a well-defined behaviour even when the throughput constraints cannot be fulfilled. Numerical results showcase the merits of our approach over the popular proportional fairness and constrained sum-rate maximization strategies.
\end{abstract}


\section{Introduction}
The ever-growing demand for fast and ubiquitous wireless connectivity poses the challenge of delivering high data rates while efficiently managing the scarce radio resources. \ac{OFDM} has become a mainstream transmission method for broadband wireless systems. Additionally, thanks to the independent fading of users' channels, efficient spectrum utilization can be attained by exploiting multiuser diversity. Specifically, \ac{OFDM} subcarriers can dynamically be allocated to multiple users according to their instantaneous channel conditions. Therefore, the multiuser version of \ac{OFDM}, called \ac{OFDMA}, has been recognized as a key technology for next-generation wireless systems.

Towards this direction, the new IEEE 802.11ax (11ax) amendment for high efficiency \ac{WLANs} employes \ac{OFDMA} in both \ac{DL} and \ac{UL} directions \cite{boris1}. Specifically, \ac{DL} \ac{OFDMA} is expected to boost \ac{DL} throughput and alleviate the \ac{DL}-\ac{UL} asymmetry problem when the \ac{AP} lacks transmission opportunities compared to the stations~\cite{boris2}. The efficiency of the \ac{OFDMA} transmissions, though, mainly hinges on how the \ac{AP} selects the stations and allocates the available resources. Therefore, intelligent multi-user scheduling is crucial for attaining the best possible system performance.

The peculiarities of 11ax OFDMA implementation render the scheduling and resource allocation problem different from that in cellular networks. There are few recent works on the OFDMA scheduling problem for 11ax networks. In~\cite{Min_Padding}, the authors proposed a framework based on Lyapunov optimization to dynamically adjust the OFDMA transmission duration so that padding overhead is minimized. To do so, they assumed flat fading across the \ac{RUs} and considered fixed RU allocation in conjunction with round-robin user scheduling. The problem of joint user scheduling and RU allocation was firstly studied in \cite{UL_Schedulers1}, \cite{UL_Schedulers2}. Specifically, D. Bankov \textit{et al}. proposed a set of multiuser schedulers by formulating the \textit{unconstrained} utility maximization problem as an assignment problem for the \ac{UL}. However, they focused on maximizing the utility of instantaneous station rates rather than their long-term throughput. In general, to provide fairness as well as minimum quality-of-service among the stations, many works in the literature proposed to maximize a utility function of the long-term time average rates under minimum average rate requirements (see~\cite{Wang}, \cite{Giannakis}, and references therein). 

In this work, we address the throughput-constrained scheduling problem in \ac{OFDMA} of 11ax. Specifically, we propound a simple yet effective scheduling policy that applies max-min fairness in order to maximize the minimum fraction between the achievable and minimum requested throughputs. Thus, it minimizes the maximum throughput violation for infeasible instances. Numerical simulations show that the propounded policy increases substantially the throughput of the worst-case station, whilst scaling efficiently as the number of stations increases in the network, with respect to existing methods such as proportional fairness and constrained sum-rate maximization. 

\textit{Notation}: $\mathcal{A}$ is a set; $\mathbf{a}$ is a vector; $(\cdot)^+ = \max(\cdot,0)$; $\max(\mathcal{A})$ denotes the maximum element of set $\mathcal{A}$; and $(\mathbf{A})_{i,j}$ is the $(i,j)$th entry of matrix $\mathbf{A}$.
\begin{figure}[t]
	\centering
	\includegraphics[scale=0.95]{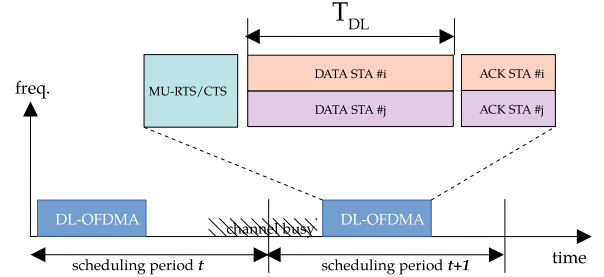}
	\caption{Illustration of a \ac{DL} OFDMA transmission.}
	\label{Fig:fig1}
\end{figure}
\section{System Model}

\subsection{\ac{DL} \ac{OFDMA} Model}
Consider the \ac{DL} of a 11ax network consisting of an \ac{AP} and a set $\mathcal{K} \triangleq \{1,\dots, K\}$ of stations. We assume a throughput-constrained traffic model\footnote{This is suitable for elastic applications, such as file transfers and Web browsing sessions~\cite{elastic_traffic_wns, unified_approach_qos}.} where each station $k$ has a minimum \ac{DL} throughput requirement denoted by $r^{\min}_k$. To this end, the \ac{AP} periodically commences a \ac{DL} \ac{OFDMA} transmission of duration $T_{\textsc{dl}}$, as shown in Fig.~\ref{Fig:fig1}. Specifically, the time axis is divided into scheduling periods of equal duration, with period~$t$ corresponding to the normalized time interval $[t, t+1)$. In each scheduling period, an OFDMA transmission takes place without collisions; if there is an ongoing transmission, the scheduled \ac{OFDMA} transmission is deferred until the channel is sensed idle. Let $\mathcal{N}\triangleq\{1,\dots,N\}$ be the set of \ac{RUs}, with each RU comprising of multiple consecutive subcarriers.\footnote{11ax supports various RU configurations. We focus on a specific configuration for ease of exposition. In the simulation results, we investigate the case of multiple RU patterns as well.} The vector of channel gains of station $k$ over RU $n$ in period $t$ is denoted by $\mathbf{g}_{k,n}[t]$. The \textit{channel state} in scheduling period $t$ is then defined as $\mathcal{G}[t]\triangleq  \{\mathbf{g}_{k,n}[t]\}_{k\in\mathcal{K}, n\in\mathcal{N}}$, and is assumed to evolve according to a block-fading process. Thus, $\mathcal{G}[t]$ remains constant during period $t$ but is independent and identically distributed across different scheduling periods. 

\subsection{Scheduling Policy}
Let $p_{k,n}[t]$ denote the transmit power assigned to the $k$th station over RU $n$ in scheduling period $t$. For simplicity, we assume equal power allocation across the $N$ RUs. Hence, $p_{k,n}[t] = P_{\text{total}}/N$, where $P_{\text{total}}$ is the power budget of the \ac{AP}. The scheduling decisions are then specified by the binary variables $s_{k,n}[t]\in\{0,1\}$, with $s_{k,n}[t] =1$ if RU $n$ is assigned to the $k$th station, and  $s_{k,n}[t] = 0$ otherwise. A scheduling policy controls the decisions of the \ac{AP} at each period $t$, which are given by the matrix $\mathbf{S}[t]$ defined as $(\mathbf{S}[t])_{k,n}\triangleq s_{k,n}[t]$. More particularly, at the beginning of each period $t$, the \ac{AP} observes the random channel state $\mathcal{G}[t]$, and selects $\mathbf{S}[t]$ whose elements satisfy the 11ax RU allocation constraints
\begin{align}
\sum_{k=1}^K s_{k,n}[t]&\leq 1, \ \forall n\in\mathcal{N}, \label{eq:ru_constraint1}\\
\sum_{n=1}^N s_{k,n}[t] &\leq 1, \ \forall k\in\mathcal{K}. \label{eq:ru_constraint2}
\end{align}
Constraint~\eqref{eq:ru_constraint1} ensures that stations cannot share the same RU, while constraint~\eqref{eq:ru_constraint2} guarantees that every station is assigned to one RU at most. Finally, the set of all feasible scheduling decisions is defined as
\begin{equation}
 \mathcal{S}\triangleq\left\{\mathbf{S}\in\{0,1\}^{K\times N} \ | \  \eqref{eq:ru_constraint1}-\eqref{eq:ru_constraint2}\right\}.
 \end{equation}

\subsection{Rate Allocation and Throughtput} For a given channel realization $\mathbf{g}_{k,n}[t]$ and transmit power $p_{k,n}[t]$, the \ac{AP} can transmit $f\left(p_{k,n}[t],\mathbf{g}_{k,n}[t]\right)$ bits per OFDM symbol to station $k$ over RU $n$. The function $f(\cdot,\cdot)$ models the rate selection scheme, and has to conform with the 11ax restriction that a single \ac{MCS} is employed over all the subcarriers of a RU \cite{802.11ax_Draft}. Next, assume there are $L$ MCSs, and let $\rho_l$ denote the bit rate of \ac{MCS} $l$. If RU $n$ consists of $S_n$ data subcarriers and all subcarriers are used for transmission, then the set of achievable bit rates on RU $n\in\mathcal{N}$ is given by $\mathcal{R}_n = \{S_n\rho_1,\dots,S_n\rho_L\}$. The number of bits transmitted to station $k$ over RU $n$ during the scheduling period $t$ is denoted by $r_{k,n}[t]$, and is calculated~as
\begin{equation}\label{eq:tx_rate_formula}
r_{k,n}[t] = f\left(p_{k,n}[t],\mathbf{g}_{k,n}[t]\right)\frac{T_{\textsc{dl}}}{T_{\textsc{ofdm}}},
\end{equation}
where $T_{\textsc{ofdm}}$ is the duration of an OFDM symbol, and $f(p_{k,n}[t],\mathbf{g}_{k,n}[t])\in\mathcal{R}_n$. Then, the instantaneous transmission rate $r_k[t]$ associated with the $k$th station is given by 
\begin{align}
r_k[t]  = \sum_{n=1}^Ns_{k,n}[t]r_{k,n}[t]  \quad (\text{bits/period}).
\end{align}
Finally, the \textit{throughput} of station $k$ is defined as the long-term time average
\begin{equation}\label{eq:longterm_rate}
\bar{r}_k \triangleq \lim_{T\to\infty}\sup\frac{1}{T}\sum_{t=0}^{T-1}\mathbb{E}\{r_k[t]\}, 
\end{equation}
where the expectation is taken with respect to the channel realizations as well as the scheduling decisions.

\section{Throughput-Constrained Scheduling}
\subsection{Problem Description}
It is customary to assess the efficacy of a scheduling policy in terms of fairness. This is modeled by a utility function $U(\cdot)$, which is a concave, continuous and entrywise non-decreasing function of the stations' throughput. Let $\bar{\mathbf{r}} = (\bar{r}_1,\dots, \bar{r}_K)$. The throughput-constrained scheduling problem is then formulated as the utility maximization problem~\cite{unified_approach_qos}
\begin{equation}\label{eq:original_problem}
\begin{matrix}
\underset{\{\mathbf{S}[t]\in\mathcal{S}\}}{\text{maximize}} & U(\bar{\mathbf{r}}) & \text{s.t.} & \bar{r}_k\geq r_k^{\min}, \ \forall k\in\mathcal{K}.
\end{matrix}
\end{equation}
\textbf{Infeasible Instances and Admission Control}: Even though the aforementioned problem can be solved using tools in stochastic network utility optimization, there is no guarantee that~\eqref{eq:original_problem} is feasible, i.e., the minimum throughput requirements will be fulfilled. This situation can occur when some stations have too week channel conditions, and/or too high throughput requirements. Traditional approaches assume that~\eqref{eq:original_problem} has a feasible solution, but it might not yield a good performance for infeasible instances. One can deal with the infeasible instances by employing admission control. However, it is not trivial to identify which station to remove from the system in order to turn~\eqref{eq:original_problem} into a feasible problem. 

\subsection{Proposed Policy}
We present a low-complexity solution to the throughput-constrained scheduling problem~\eqref{eq:original_problem}, which does not depend on admission control and has a well-defined behaviour for infeasible instances. First, let $\bar{\mathbf{r}}^{\pi}$ denote the throughput vector attained by scheduling policy $\pi$. Then,~\eqref{eq:original_problem} is feasible whenever a policy $\pi$ exists with $\bar{r}_k^{\pi}/r_k^{\min} \geq 1, \forall k\in\mathcal{K}$. Therefore, we propose to directly maximize the minimum $\bar{r}_k/r_k^{\min}$, leading to the optimization problem
\begin{equation}\label{eq:wmaxmin_problem}
\begin{matrix*}[l]
& \underset{\{\mathbf{S}[t]\in\mathcal{S}\}}{\text{maximize}}              & \underset{k\in\mathcal{K}}{\min} \left\{\mathlarger{\frac{\bar{r}_k}{r_k^{\min}}}\right\}.
\end{matrix*}
\end{equation}
Note that the weighted max-min problem~\eqref{eq:wmaxmin_problem} admits a well-defined solution. More particularly, if $\bar{r}_k/r_k^{\min} \geq 1,  k\in\mathcal{K}$, then the policy that solves~\eqref{eq:wmaxmin_problem} maximizes the minimum surplus $\bar{r}_k - r_k^{\min}\geq 0$. Likewise, if there is at least one station $k'\in\mathcal{K}$ with $\bar{r}_{k'}/r_{k'}^{\min} < 1$, the policy that solves~\eqref{eq:wmaxmin_problem} minimizes the maximum constraint violation $r^{\min}_{k'} - \bar{r}_{k'}~\geq 0$.

\section{Solution via Lyapunov Optimization}
In the sequel, we resort to Lyapunov optimization to derive a near-optimal solution.
\begin{definition}
Let $U^{\star}$ denote the maximum utility value of~\eqref{eq:wmaxmin_problem}, and  $\bar{\tilde{\mathbf{r}}} \triangleq (\bar{r}_1/r_1^{\min}, \dots, \bar{r}_K/r_K^{\min})$. A scheduling policy $\pi$ is said to produce an $O(\epsilon)$-optimal solution to~\eqref{eq:wmaxmin_problem} if $U\left(\bar{\tilde{\mathbf{r}}}^{\pi}\right) \geq U^{\star} - O(\epsilon)$, and all the associated constraints are satisfied.
\end{definition}
\subsection{The Transformed Problem}\label{sec:transformed_problem}
Following the approach in~\cite[Ch. 5]{SNO1} for solving stochastic network optimization problems, we transform the problem~\eqref{eq:wmaxmin_problem} into a form involving only time averages rather than functions of time averages. To this end, let $\bm{\gamma}[t] = (\gamma_1[t],\dots,\gamma_K[t])$ be a vector of auxiliary variables chosen within a set $\Gamma$. The set $\Gamma$ must bound both the auxiliary and rate variables, and hence is selected as
\begin{equation}
\Gamma = \left\{\bm{\gamma}\in\mathbb{R}^K \ | \  0 \leq \gamma_k \leq R_{\max}, \ \forall k\in\mathcal{K}\right\},
\end{equation}
where $R_{\max} = \max(\bigcup_{n=1}^N\mathcal{R}_n)$ is the maximum transmission rate over a RU. Now consider the transformed problem:
\begin{equation}\label{eq:transformed_problem_maxmin}
\begin{matrix*}[l]
 \underset{\{\mathbf{S}[t]\in\mathcal{S}\}, \{\bm{\gamma}[t]\in\Gamma\}}{\text{maximize}}                 & \overline{\underset{k\in\mathcal{K}}{\min}\{\gamma_k[t]\}} &\\\\
 \ \ \ \ \ \ \  \text{s.t.}                  & \bar{\gamma}_k \leq \bar{r}_k/r_k^{\min},  & \forall k\in\mathcal{K}.
\end{matrix*}
\end{equation} 
The connection between \eqref{eq:wmaxmin_problem} and \eqref{eq:transformed_problem_maxmin} is established as follows. Suppose an arbitrary scheduling policy $\pi$ solves the problem~\eqref{eq:transformed_problem_maxmin}. The maximum utility value, denoted by $\overline{U(\bm{\gamma}^{\pi})}$, is then attained, whilst satisfying all the associated constraints. Because $U(\cdot)$ is concave, it also holds
\begin{equation}\label{eq:inequalities}
\bar{\tilde{\mathbf{r}}}^{\pi} \geq \bar{\bm{\gamma}}^{\pi} \Rightarrow U\left(\bar{\tilde{\mathbf{r}}}^{\pi}\right) \geq U\left(\bar{\bm{\gamma}}^{\pi}\right) \geq \overline{U\left(\bm{\gamma}^{\pi}\right)},
\end{equation}
where the last inequality is Jensen's inequality for concave functions. According to~\eqref{eq:inequalities}, if $\overline{U\left(\bm{\gamma}^{\pi}\right)} \geq U^{\star} - O(\epsilon)$, then $U\left(\bar{\tilde{\mathbf{r}}}^{\pi}\right) \geq U^{\star} - O(\epsilon)$ as well, hence yielding the desired result. Next, we detail the algorithm the produces such a solution.

\begin{algorithm}[t]
	\caption{Weighted Max-Min Scheduling}
	
	\begin{algorithmic}[1]
		\Statex For each scheduling period $t\in\{0,1,2,\dots\}$ do:
		\State Observe $\{Q_k[t]\}_{k\in\mathcal{K}}$ and the channel state $\mathcal{G}[t]$.
		\State Choose $\bm{\gamma}(t)\in\Gamma$ such that
		\[
		\text{maximize} \ \ V\min_{k\in\mathcal{K}}\{\gamma_k[t]\}- \sum_{k=1}^K Q_k[t]\gamma_k[t].
		\]
		\State Choose $\mathbf{S}[t]\in\mathcal{S}$ such that
		\begin{align*}
		\begin{matrix*}[l]
		\text{maximize}  & \sum_{k=1}^K Q_k[t] \frac{r_k[t]}{r_k^{\min}}.
		\end{matrix*}
		\end{align*}
		\State Update the virtual queues using \eqref{eq:virtual_queue_pc}.
	\end{algorithmic}\label{algo}
	\addtocounter{algorithm}{-1}
\end{algorithm}
\subsection{The Drift-Plus-Penalty Algorithm}
In Lyapunov optimization, each time average constraint is associated with a virtual queue, and constraint satisfaction is expressed as a queue stability problem. More particularly, for the constraint $\bar{\gamma}_k \leq \bar{r}_k/r_k^{\min}$, we consider a virtual queue that obeys the recursion 
\begin{equation}\label{eq:virtual_queue_pc}
Q_k[t+1] = \left(Q_k[t] - r_k[t]/r_k^{\min} + \gamma_k[t]\right)^+,
\end{equation}
where $Q[t]$, $r_k[t]/r_k^{\min}$, and $\gamma_k[t]$ correspond to the virtual queue size, service rate, and arrival rate for the scheduling period $t$, respectively. The drift-plus-penalty (DPP) algorithm is described in Algorithm~\ref{algo}; the parameter $V$ affects the convergence speed and accuracy of the algorithm~\cite{SNO1}. 

\subsection{Maximization Subproblems}
According to Algorithm 1, we can address the weighted max-min problem by solving a set of deterministic subproblems at every scheduling period~$t$. More specifically, the first subproblem regards the auxiliary viarables. The optimal auxiliary variables are obtained by observing that
\begin{align}\label{eq:aux_ineq}
\max_{\bm{\gamma}[t]\in\Gamma} \ V\min_{k\in\mathcal{K}}\{\gamma_k[t]\} - \sum_{k=1}^KQ_k[t]\gamma_k[t]  \nonumber\\
\leq \max_{\bm{\gamma}[t]\in\Gamma} \ \gamma_{\min}[t] \left(V-\sum_{k=1}^KQ_k[t]\right),
\end{align}
\\
where $\gamma_{\min}[t] = \min_{k\in\mathcal{K}}\{\gamma_k[t]\}$. Based on~\eqref{eq:aux_ineq}, it is straightforward to show that 
\begin{equation*}
\gamma^{\star}_{k}[t] = \left\{
\begin{array}{ll}
R_{\max}, & \text{if} \ V > \sum_{k=1}^KQ_k[t] \\ \\
0, &\text{otherwise}\\ 
\end{array}
\right. .
\end{equation*}
Next, the maximization subproblem for the scheduling decision $\mathcal{S}[t]$ is recast as
\begin{equation}\label{eq:insta_max}
\begin{aligned}
&\underset{\mathbf{S}[t]\in\mathcal{S}}{\text{maximize}}
&& \sum_{k=1}^K\sum_{n=1}^Ns_{k,n}[t]\phi_{k,n}[t],
\end{aligned}
\end{equation}
where $\phi_{k,n}[t] \triangleq Q_k[t] r_{k,n}[t]/r_k^{\min}$. The transmission rate $r_{k,n}[t]$ is calculated usinig~\eqref{eq:tx_rate_formula}. The above problem determines the optimal RU assignment for the scheduling period $t$, and is a classical assignment problem. Hence, it can be solved in $O(\max(K,N)^3)$ using the Hungarian method~\cite{hungarian}. 

\section{Performance Evaluation}
In this section, we assess the performance of the proposed scheduling policy. For this purpose, we consider the following benchmark strategies:
\begin{itemize}
	\item {Proportional Fairness (PF):} In each period $t$, the scheduler selects the stations that maximize the instantaneous weighted sum-rate, where the weight associated with station $k$ is equal to the inverse of the exponential moving average of its throughput~\cite{UL_Schedulers2}.
	\item {Ergodic Sum-Rate Maximization (ESRM):} We solve the problem in~\eqref{eq:original_problem} for $U(\bar{\mathbf{r}}) = \sum_{k=1}^K \bar{r}_k$. Since $U(\bar{\mathbf{r}})$ is a linear function of the rates in ESRM, we have $U(\bar{\mathbf{r}})=\overline{U(\mathbf{r})}$. Therefore, we can readily employ the DPP algorithm without using auxiliary variables~\cite{SNO1}. To this end, we have a virtual queue for the minimum throughput constraint $r^{\min}_k\geq\bar{r}_k$, which obeys the recursion
	\begin{equation}
	Z_k[t+1]  = \left(Z_k[t] - r_k[t] + r_k^{\min}[t]\right)^+, \ \forall k\in\mathcal{K}.
	\end{equation}
	Then, in each period $t$, the scheduler selects the stations by solving~\eqref{eq:insta_max} for 	
	\begin{equation}
	\phi_{k,n}[t] = V_{\text{ESR}} r_{k,n}[t] + Z_k[t](r_{k,n}[t] - r_k^{\min}),
	\end{equation}
	where $V_{\text{ESR}}$ denotes the control parameter of the DPP algorithm for the ESRM scheme. 
\end{itemize} 
\begin{figure*}
	\centering
	\begin{subfigure}{0.85\columnwidth}
		\includegraphics[width=\columnwidth]{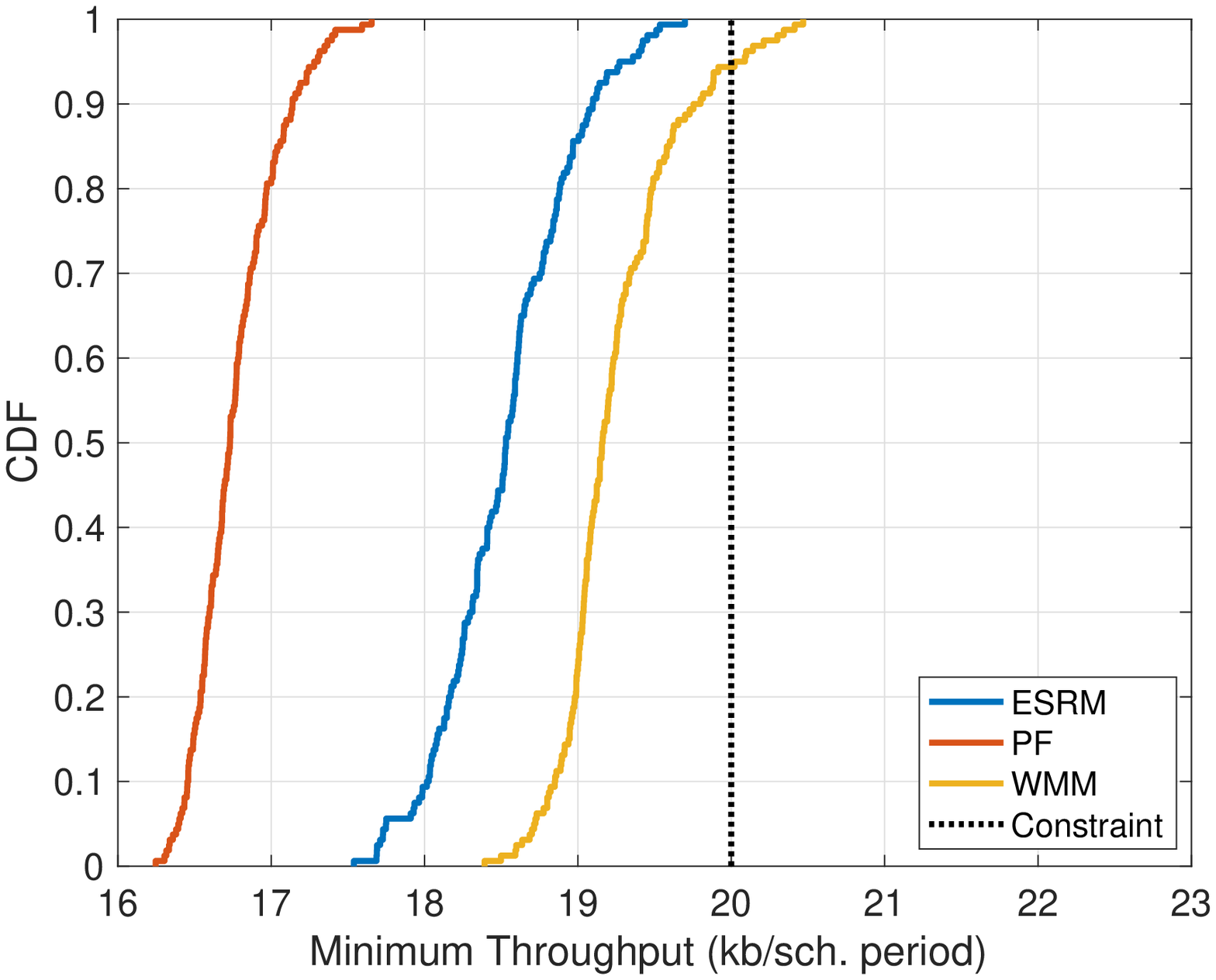}
		\caption{}
		\label{Fig:fig2a}
	\end{subfigure}\hfil
	\begin{subfigure}{0.85\columnwidth}
		\includegraphics[width=\columnwidth]{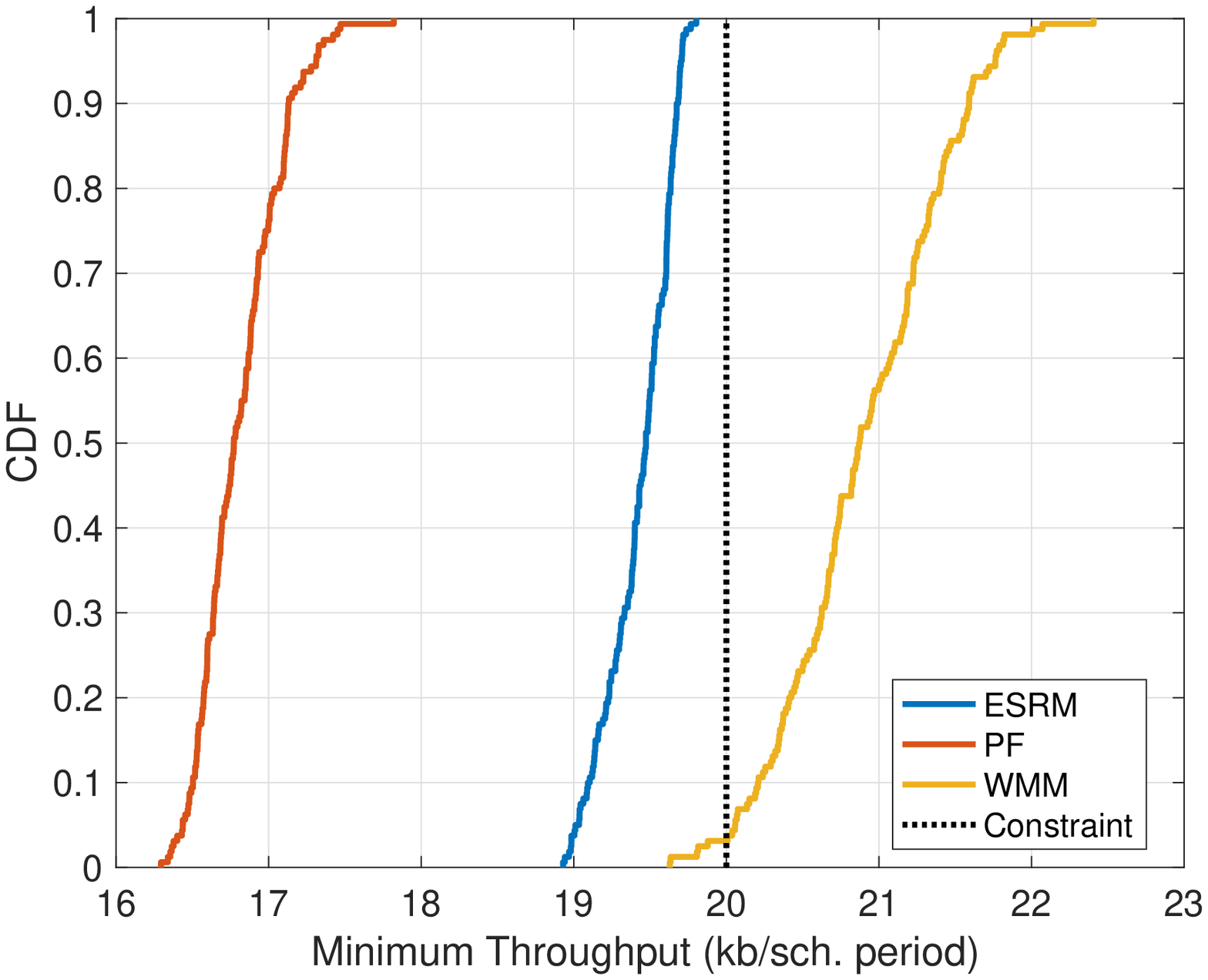}
		\caption{}
		\label{Fig:fig2b}
	\end{subfigure}
	\caption{Empirical CDF of the minimum throughput for $K=12$ stations: (a) single RU pattern $(N_1, S_1)$; (b) multiple RU patterns.}
	\label{Fig:fig2}
\end{figure*}
\begin{table}[H]
	\centering
	\caption{Main simulation parameters.}
	\label{Table:sim_params}
	\begin{tabular}{| l | l | l |} 
		\hline
		\textbf{Notation} & \textbf{Description}  & \textbf{Value}\\ 
		\hline
		$f_c$ & Carrier frequency & $5$ GHz\\
		$d_{\max}$ & Radius of the WLAN area  & $15$ m  \\
		$P_{\text{total}}$ & Maximum transmit power & $20$ dBm \\
		$T_{\textsc{ofdm}}$ & Duration of OFDM symbol & $16$ $\mu\text{s}$ \\
		$T_{\textsc{dl}}$ & Duration of \ac{DL} OFDMA transmission & $3.2$ $\text{ms}$ \\
		\hline
	\end{tabular}
\end{table}
\begin{table}[H]
	\centering
	\caption{\ac{MCS} for the 20 MHz channel~\cite{802.11ax_Draft}.}
	\begin{tabular}{|c | c | c |} 
		\hline
		\textbf{Index} & \textbf{MCS} & \textbf{Minimum SNR (dBm)} \\ 
		\hline
		1 &BPSK, $1/2$  		 & $-82$\\
		2 &QPSK, $1/2$  		& $-79$\\
		3 &QPSK, $3/4$  		& $-77$\\
		4 &16-QAM, $1/2$    & $-74$\\
		5 &16-QAM, $3/4$    & $-70$\\
		6 &64-QAM, $2/3$    & $-66$\\
		7 &64-QAM, $3/4$    & $-65$\\
		8 &64-QAM, $5/6$    & $-64$\\
		9 &256-QAM, $3/4$  & $-59$\\
		10 &256-QAM, $5/6$  & $-57$\\
		\hline
	\end{tabular}
	\label{table: MCS}
\end{table}
\subsection{Simulation Setup}
The area of the WLAN is modeled by a circle of radius $d_{\max}$ meters. The \ac{AP} is located at the center of the circle, and stations are uniformly distributed inside the circle with minimum distance from the AP of $1$ meter. The path attenuation is calculated using the 11ax path-loss model for a residential scenario \cite{802.11ax_Scenarios}
\begin{align*}
\text{PL}_k = 40.05 &+ 20\log_{10}(f_c/2.4) + 20\log_{10}(\min(d_k,5)) \\ 
&+ 1\{d_k>5\}\cdot35\log_{10}(d_k/5), \quad (\text{dB})
\end{align*}
where $f_c$ is th carrier frequency in GHz, $d_k$ is the distance between the AP and the $k$th station, and $1\{\cdot\}$ denotes the indicator function. The channel bandwidth is $20$ MHz, and is divided into $N$ \ac{RUs}. Without loss of generality, we assume that each RU consists of $S$ subcarriers. We also consider Rayleigh fading across the \ac{RUs}. Let $g_{k,n}$ denote the channel gain of station $k$ over RU $n$. The power $p_{k,n}$ of station $k$ is uniformly distributed among the subcarriers of RU $n$, and therefore the received \ac{SNR} at each subcarrier is 
\[
\text{SNR}_{k,n}=  10\log_{10}\left(\frac{p_{k,n}}{S} \right) - \text{PL}_k + 10\log_{10}(g_{k,n})  \ (\text{dBm}).
\]
Based on the received \ac{SNR}, the maximal \ac{MCS} is selected, which is denoted by $l^*$. The bit rate and \ac{SNR} threshold of each \ac{MCS} are given in Table \ref{table: MCS} of the Appendix. Finally, It is worth noting that 256-QAM with $5/6$ code rate is the highest available \ac{MCS} for 26-tone RUs; 1024-QAM is an optional feature for RUs with at least 242 tones each~\cite{802.11ax_Draft}. Next, the rate of station $k$ over RU $n$ is calculated as
\[
r_{k,n} = S \rho_{l^*} \frac{T_{\textsc{dl}}}{T_{\textsc{ofdm}}}   \quad (\text{bits}/\text{sch. period}).
\]
The values of the main simulation parameters are given in Table \ref{Table:sim_params}. For the auxiliary variables, the option set is $\Gamma = \{S \rho_1,\dots,S\rho_L\}$; $\rho_1$ is the bit rate of BPSK with code rate $1/2$, and $\rho_L$ is the bit rate of $256$-QAM with code rate~$5/6$.

\subsection{Numerical Results}

\subsubsection{Single versus Multiple RU Patterns} We first investigate the benefits of employing multiple RU patterns. We use the minimum throughput achieved by a station as a figure of merit. To this end, we evaluate the empirical \ac{CDF} of the minimum throughput for 100 network realizations. Specifically, for each network realization, the minimum throughput is calculated by averaging over 1000 small-scale fading realizations. Regarding the 11ax multiple RU patterns, we consider the following three options:
\begin{itemize}
	\item $N_1 = 9$ RUs with $S_1 = 24$ data subcarriers each. 
	\item $N_2 = 4$ RUs with $S_2 = 48$ data subcarriers each.
	\item $N_3 = 2$ RUs with $S_3 = 102$ data subcarriers each. 	
\end{itemize}
We then solve~\eqref{eq:insta_max} for each pattern under equal power allocation across RUs, and choose the one that yields the maximum objective value. We plot the results in Fig.~\ref{Fig:fig2} for a minimum throughput requirement $r_k^{\min} = 20$ kb/sch. period, $\forall k\in\mathcal{K}$. The control parameters of the DPP algorithm for the WMM and ESMR are $V_{\text{WMM}} = 900$ and $V_{\text{ESRM}} = 10$, respectively. As we observe, the WMM outperfoms both the PF and ESRM policies. Note that, although the use of multiple RU patterns might increase the computational complexity of the scheduler, it yields substantial performance gains for the WMM policy. In particular, the minimum throughput requirement $r^{\min}$ corresponds to the $95$th percentile when a single RU pattern is employed, whilst it drops to the $3$th percentile for multiple RU patterns. 

\subsubsection{Number of Stations}
\begin{figure*}
	\centering
	\begin{subfigure}{0.85\columnwidth}
		\includegraphics[width=\columnwidth]{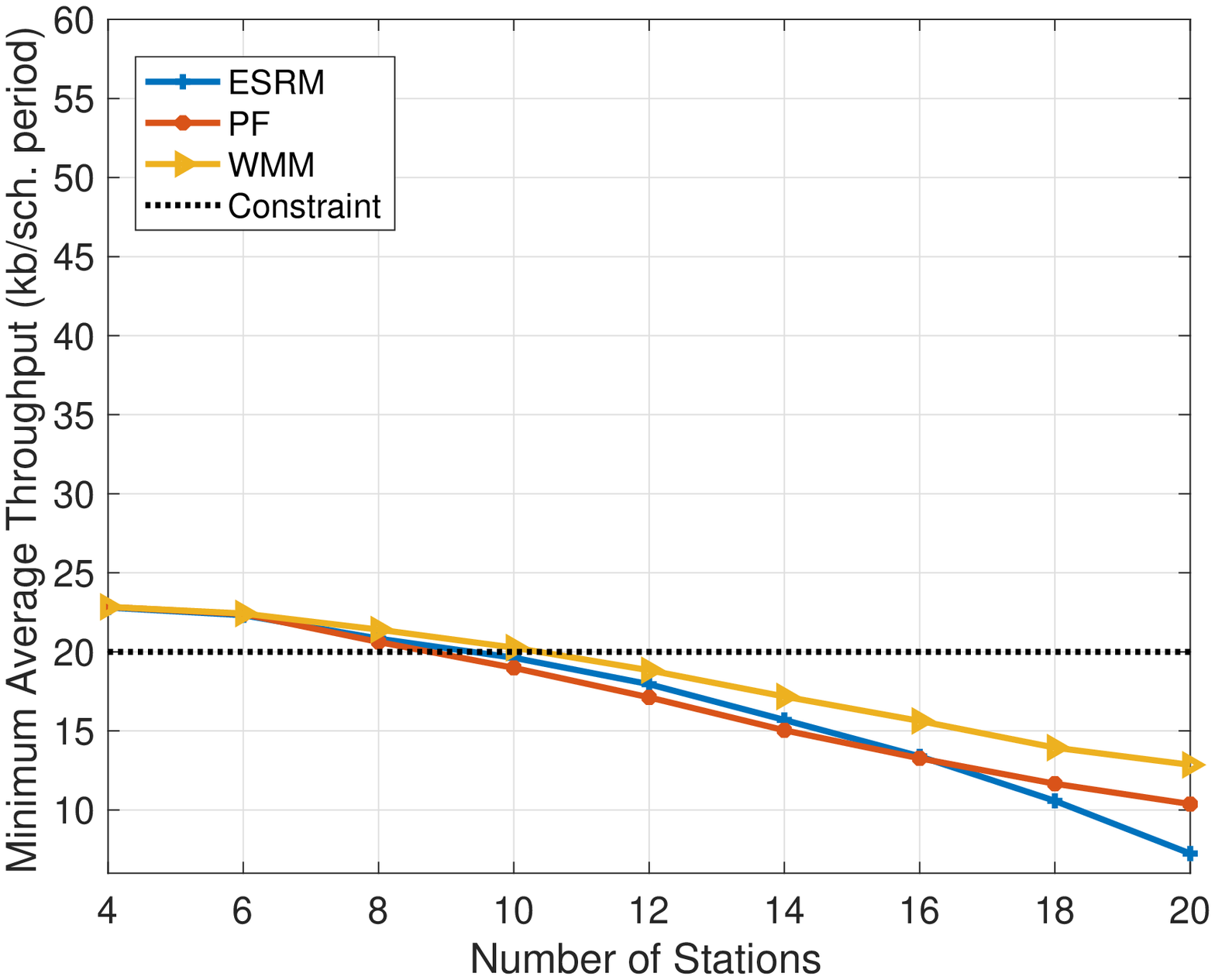}
		\caption{}
		\label{Fig:fig3a}
	\end{subfigure}\hfil
	\begin{subfigure}{0.85\columnwidth}
		\includegraphics[width=\columnwidth]{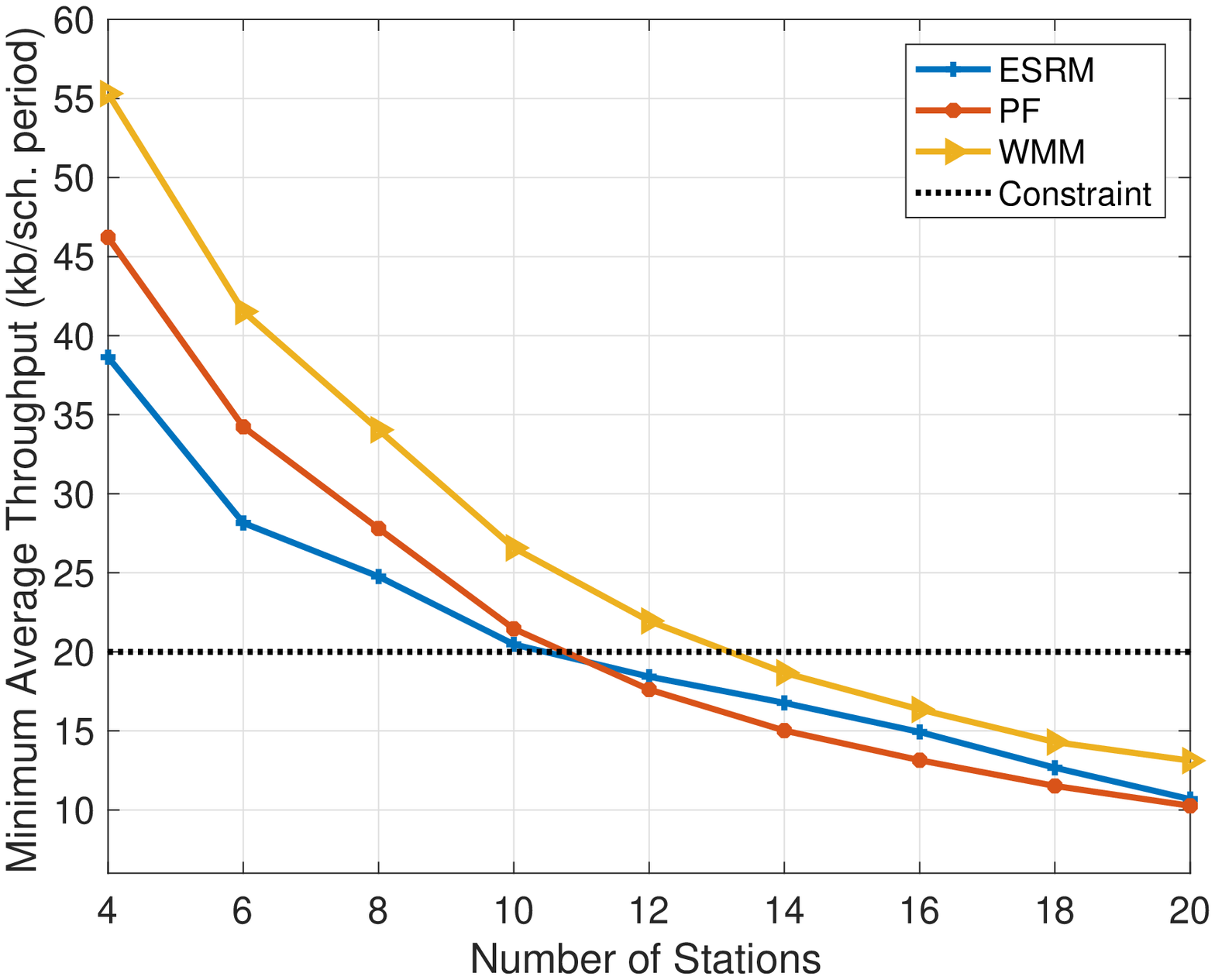}
		\caption{}
		\label{Fig:fig3b}
	\end{subfigure}
	\caption{Minimum average throughput as a function of the number of stations: (a) single RU pattern $(N_1, S_1)$; (b) multiple RU patterns.}
	\label{Fig:fig3}
\end{figure*}

We now examine how the performance of the schedulers under consideration scales as the number of stations increases. To this end, we calculate the average of the minimum throughput for the same number of network and small-scale fading realizations as previously. From Fig.~\ref{Fig:fig3}, we see that the WMM delivers the highest minimum throughput even when the throughput constraint of $20$ kb/sch. period cannot be fulfilled. Consequently, the performance of the proposed scheduler scales more efficiently compared to the PF and ESRM schedulers. Finally, we stress that the WMM policy benefits the most from the utilization of multiple RU patterns. 
\section{Conclusions}
In this letter, we provided a solution to the DL OFDMA scheduling problem in 11ax with minimum throughput requirements. First, we introduced a scheduling and rate allocation model which conforms with the 11ax implementation constraints. Then, we formulated the throughput-constrained scheduling problem as an unconstrained problem using weighted max-min fairness. Relying on Lyapunov optimization, we derived a dynamic scheduling policy which minimizes the maximum throughput constraint violation for infeasible instances, and maximizes the minimum throughput surplus otherwise. We finally provided numerical results showcasing that our approach outperforms the popular proportional fairness and constrained sum-rate maximization strategies in terms of the minimum throughput.

\end{document}